\documentclass[reprint,amsmath,amssymb,apr,aip,onecolumn, 10pt]{revtex4-1}
\usepackage{graphicx}
\usepackage{graphics}
\usepackage{mathptmx}
\usepackage{times}
\usepackage{amsmath}
\usepackage{amssymb}
\usepackage{dcolumn}
\usepackage{bm}
\usepackage{units}
\usepackage{color}
\usepackage{soul}
\usepackage{xcolor}

\renewcommand{\thesubsection}{\arabic{section}.\arabic{subsection}}

\usepackage{lineno,hyperref}
\begin{document}






\title{Bio-mimetic Synaptic Plasticity and Learning in a sub-500mV Cu/SiO$_2$/W Memristor}
\author{S. R. Nandakumar}  \affiliation{New Jersey Institute of Technology, Newark, NJ 07102, USA}
\author{Bipin Rajendran}\email{bipin.rajendran@kcl.ac.uk} \affiliation{King's College London, Strand, London WC2R 2LS, UK}
\date{\today}

\begin{abstract}
The computational efficiency of the human brain is believed to stem from the parallel information processing capability of neurons with integrated storage in synaptic interconnections programmed by local spike triggered learning rules such as spike timing dependent plasticity (STDP).  The extremely low operating voltages (approximately  $100\,$mV) used to trigger neuronal signaling and synaptic adaptation is believed to be a critical reason for the brain's power efficiency.  We demonstrate the feasibility of spike triggered STDP behavior in a two-terminal Cu/SiO$_2$/W memristive device capable of  operating below   $500\,$mV. We analyze the state-dependent nature of conductance updates in the device to develop a phenomenological model. Using the model, we evaluate the potential of such devices to generate precise spike times under supervised learning conditions and classify handwritten digits from the MNIST dataset in an unsupervised learning setting. The results form a promising step towards creating a low power synaptic device capable of on-chip learning.  
\end{abstract}




\maketitle

\section{Introduction}

Today's state-of-the-art computing platforms are capable of performing quadrillions of precise logical operations in a second, albeit consuming millions of Watts. However, they still lag behind the capabilities and efficiencies of the human brain, especially for unstructured data analytics and decision-making tasks. The reason behind the inefficiency of conventional machines for such applications is the underlying von Neumann architecture, requiring constant shuttling of data between the physically separated processor and memory units. In contrast, the human brain is a massively parallel information processing system with approximately 100 billion neurons that interact with each other by transmitting electrical signals through 1000 trillion neuronal junctions or synapses.  The parallel operation of networks in the brain which integrates memory locally in its synaptic junctions and their plastic nature enable adaptation and learning depending on neuronal spike based information processing activities.    
The von Neumann machines we use today are incapable of such learning and adaptation and executes its tasks based on pre-programmed algorithms in a sequential manner. This stark difference in the two computational paradigms calls for architectural innovations that could aid the development of intelligent computing platforms that could learn in real-time and in the field. Synaptic adaptation based on neuronal signaling activity is believed to be the fundamental basis of learning and memory in the brain \cite{Kandel}. Artificial neural networks with similar plastic synapses have been demonstrated to be capable of performing intelligent analytic tasks, even rivaling human performance \cite{game,SzegedyIV16}. Hence, there is significant interest in building artificial systems that mimic the key architectural features of the brain, such as the parallel event-driven communication through low-power-programmable nanoscale components.

\begin{figure}[th!]
	\centering
	\includegraphics[width=0.8\linewidth]{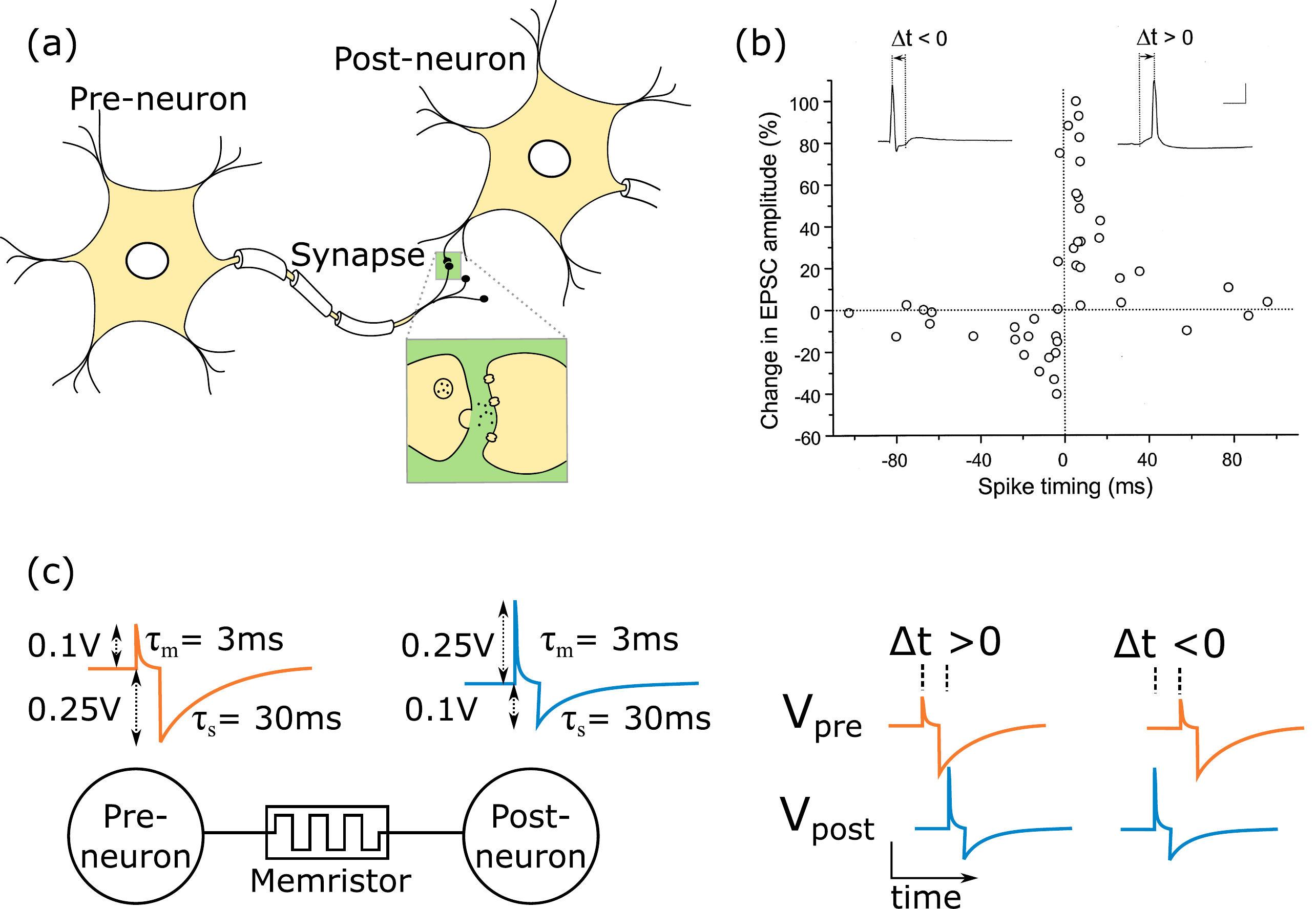}
	\caption{\textbf{Spike timing dependent plasticity} (a) Neurons connect and exchange information via junctions between axons and dendrites, called synapses. (b) Synaptic conductance modification observed in a rat hippocampal neuron\cite{BiPoo}: an example of spike timing dependent plasticity (STDP) observed in biology. Positive and negative changes in excitatory post-synaptic current (EPSC) indicate synaptic potentiation and depression, respectively. (c) Bio-inspired programming waveforms used in the experiment to demonstrate STDP in a memristive device.}
	\label{fig1}
\end{figure}

The neurons in the brain encode information by issuing voltage signals called action potentials or spikes whose amplitude is around $100\,$mV above the resting potential. These spikes are then transmitted in a parallel fashion along the axons to thousands of other neurons through synaptic junctions {(Fig.\ref{fig1}a)}. The post-synaptic neurons issue further spikes based on the integration of weighted sum of its inputs. Coincident spiking activity of the pre- and post-synaptic neurons alters the effective conductivity of the synapses \cite{Nelson2001, Citri2007} due to the insertion or internalization of receptor molecules on the synaptic terminals. 
Such unsupervised local training rules are believed to be integral to the energy and computational efficiency of the brain \cite{Zenke2015, Abbott04, Nessler2013, Masquelier08}.  An example of a spike timing dependent plasticity (STDP) rule observed in a rat hippocampal neuron is shown in Fig.\ref{fig1}b \cite{BiPoo}, where a post-neuron spike immediately following a pre-neuron spike (within 40--80\,ms) leads to an increase in conductance (potentiation) and a pre-neuron spike following a post-neuron spike leads to a decrease in conductance (depression). {Resistive memory devices, also known as memristors, have been shown to be capable of similar conductance modulations  as in a synapse  when applied with suitable electrical pulses. We copy such a preneuron-memristor-postneuron arrangement and bio-mimicking programming waveforms to realize STDP behavior in a memristive device in this work (Fig.\ref{fig1}c)}.

Mimicking human brain's architecture and computational primitives to build intelligent information processing systems is the key goal of neuromorphic engineering research activities worldwide. While there have been several demonstrations of  neuromorphic computational platforms  using standard complementary metal-oxide semiconductor (CMOS) technology \cite{truenorth, Esser2016, SpiNNaker, Neurogrid, schemmel10, Gehlhaar:ASPLOS2014, Qiao_ROLLS, Chen:JSSCC16, Bang:ISSCC17, Moons:ISSCC17}, and demonstrations of  nanoscale devices to mimic neuronal and synaptic dynamics \cite{Tuma, Jerry2016, Rajendran18, Rajendran19, Rajendran30, jackson13, querlioz2015bioinspired, kuzum12, Prezioso2015, Nandakumar2016}, none of these have achieved the target energy efficiency specifications necessary to build systems that can learn in real-time and in-the-field \cite{Rajendran13specs}.

In order to build systems that mimic the massive parallelism and local learning aspects of the human brain, compact electronic devices that implement the dynamics of neurons and synapses are required. 3D integration of crossbar circuits with synaptic devices at its junctions and neuronal devices at the periphery is an efficient architectural paradigm to achieve the parallel connectivity of the human brain, mitigating the memory-processor bottleneck of conventional von Neumann systems. Memristive devices that exhibit conductivity modulation based on past programming history are excellent candidates to realize synaptic memory \cite{Strukov2008, Mazumder2012}. There have been numerous synaptic device demonstrations in oxides \cite{Park2012a, Park2013b, Chen2016}, and chalcogenides \cite{Suri11, jackson13} which show analog conductivity modulation based on non-volatile rearrangements of atomic configurations within the active volume of the device.  However, these devices present many challenges in terms of programming stochasticity and asymmetry, granularity, reliability, and the energy required for implementing conductivity modulation, and no single device has so far achieved all the target specifications. For example, Ta/TaO$_x$/TiO$_2$/Ti device has demonstrated femto-Joule level programming energies, but requires programming voltages above $18\,$V \cite{Wang2014}. Similarly, chalcogenide-based phase change memory devices have pico-Joule level programming energies, however, the high programming current limits parallel programmability and require access transistors at every cross-point in an array\cite {jackson13}.   
Hence, physics-driven device engineering to improve various synaptic device requirements and finding the right trade-offs for the targeted applications are necessary.

Resistive random access memories (RRAMs) are non-volatile memory devices with a metal-insulator-metal structure capable of realizing synaptic networks of high integration density. These devices store information in their resistance states which could be modulated by external programming pulses. The resistance transition in these devices typically involves a redox reaction where ionized atoms migrate in the direction of the field and get reduced at one of the electrodes.  Depending on the nature of the ionic species involved in the atomic rearrangement, these devices are classified either as Oxide-RAMs (OxRAMs) or conductance bridge-RAMs (CBRAMs). The conductance modulation in OxRAMs is due to the movement of Oxygen atoms or their vacancies in the dielectric, while that in CBRAMs is due to the movement of metallic ions (eg. Cu, Ag) from one of the electrodes.
Due to the relatively higher mobility of the metallic ions in dielectrics such as SiO$_x$ and GeS$_x$, CBRAMs tend to have smaller operating voltages compared to OxRAMs.  The non-uniformities or the defects in the thin dielectric lead to localization of the applied programming field and hence the atomic rearrangement often leads to the formation of low resistance nano-filamentary structures. Further, the relatively higher resistivity contrast between the dielectric and the filament material in CBRAM accentuates the electric field in the filament gap which accelerates the filament growth/decay and conductance switching. While the low operating voltages of the CBRAM is advantageous, a more gradual conductance change is desirable for synaptic implementation. Doping the dielectric with the active electrode material (eg. Cu) could reduce the resistivity contrast between the dielectric and filament material and could improve the probability of a non-filamentary conductance transition. Doping the dielectric by annealing has been demonstrated to achieve gradual conductance change in Cu/SiO$_x$/W based device\cite{Chen2016}. Also, the feasibility of using CBRAM devices for STDP has been shown by model simulations\cite{Yu2010b}. 

In this work, we experimentally demonstrate STDP in a Cu/SiO$_2$/W based memristor. Using a computationally simpler phenomenological model of the device that captures all its salient operating characteristics, we demonstrate the feasibility of using these devices in both supervised and unsupervised learning scenarios. We also evaluate the learning performance of these netwrorks with non-ideal device characteristics and estimate  the synaptic energy consumption under chosen programming scheme. 

\color{black}

\section{Results}
\subsection{Synaptic device based on Cu/SiO$_2$/W memristor}

Our device has a $10\,$nm SiO$_2$ dielectric layer sputter deposited between a Cu top electrode (TE) and a W bottom electrode (BE). The device is fabricated as a cross-point structure using a two mask process.  The SiO$_2$ dielectric and Cu electrode are deposited without breaking the vacuum and are patterned together. The device undergoes a final $400^{\circ}$C, $5\,$min anneal during which Cu species diffuses into the dielectric (Supplementary Note 1). 

\begin{figure}[!ht]
	\centering
	\includegraphics[width = 1\textwidth]{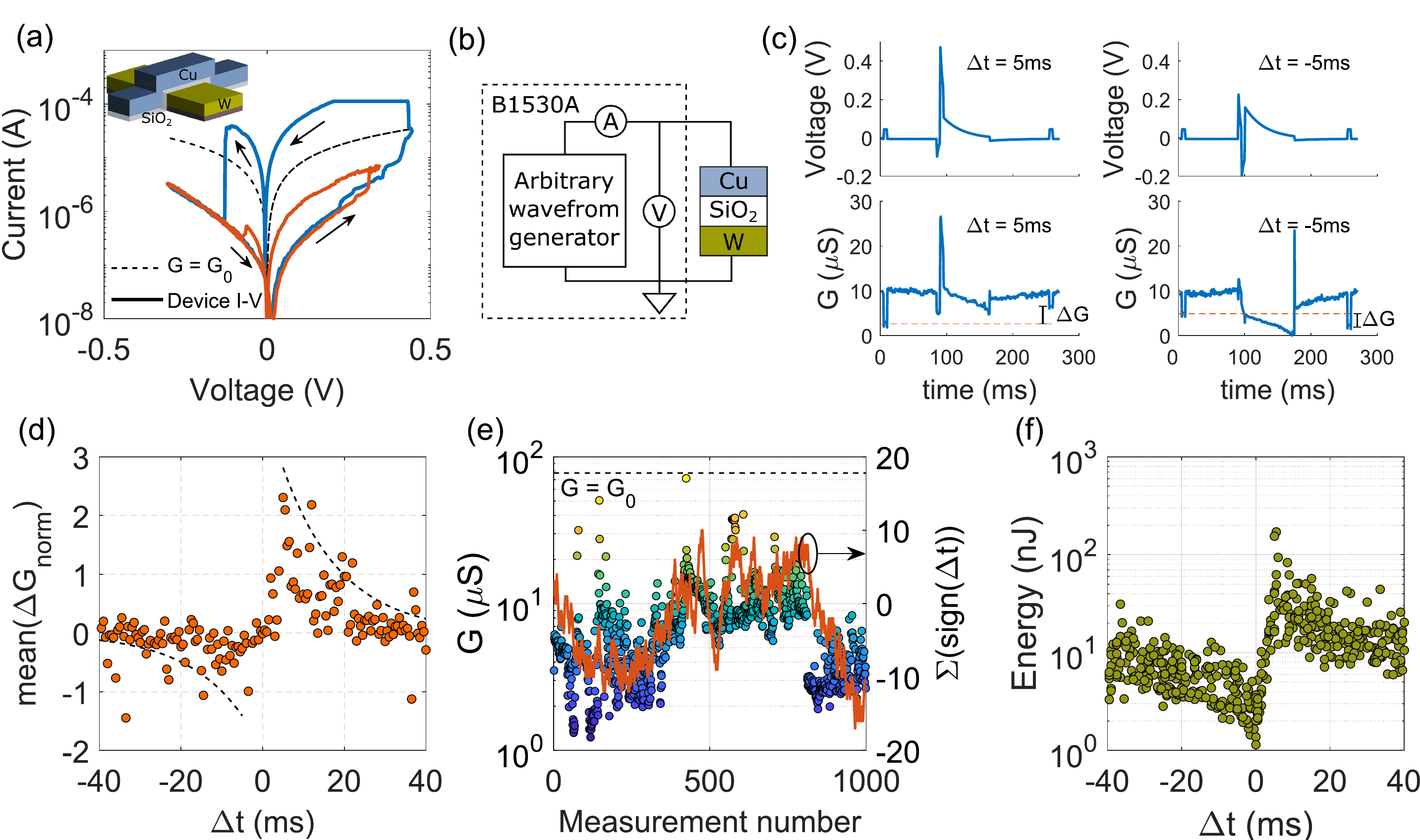}
	\caption{\textbf{STDP in memristor} (a) Memory switching behavior of the Cu/SiO$_2$/W cross-point device for two voltage sweeps with maximum amplitude of 350\,mV and 450\,mV. The current corresponding to a conductance of G$_0 (=2e^2/h)$ is marked using a dashed line. The conductance response above the G$_0$ level (blue curve) indicates at least one filamentary path connecting the top and bottom electrode, while the orange curve shows partial conductance switching in the sub-quantization regime. The 3D device structure is shown in the inset. (b) STPD programming and measurement set-up (c) Example STDP programming waveforms applied (top) and the measured device conductance evolution (bottom) where $\Delta t = 5\,$ms results in  a synaptic potentiation and $\Delta t = -5\,$ms results in a synaptic depression. (d) STDP response of the device determined as the average conductance change (normalized) versus the spike time difference based on 400 measurements. $\Delta G_{norm} = (G_f-G_i)/min(G_i, G_f)$. {The black dotted line is an eye guide approximately indicating the maximum conductance change (normalized) for each $\Delta t$.} (e) Evolution of the device conductance during the STDP measurement tracks the overall causal/anti-causal signal correlation. The orange curve is a cumulative sum of the sign of the $\Delta t$s across the sequence of measurements. (f) Energy consumption per spike pair as a function of  $\Delta t$ for the chosen programming waveforms and the conductance range observed from the device.}
	\label{fig2}
\end{figure}

The device exhibits typical memristive pinched hysteresis I-V with switching threshold below $500\,$mV (Fig.\ref{fig2}a). Bipolar voltage sweeps of approximately 2V/s ramp rate are applied to the Cu electrode while the W is kept at ground potential. 
A sufficiently large positive voltage across the device results in the ionization of relatively active Cu atoms. The migration of the resulting ions under the electric field and reduction at the bottom electrode eventually leads to the formation of metallic nano-filament paths connecting the top and bottom electrode causing the device to switch from a high resistance state (HRS) to a low resistance state (LRS), known as SET transition. Previously, we observed conductance quantization as half-integer multiples of quantum conductance $G_0=2e^2/h$ in its LRS state during current sweep measurements confirming the presence of nano-filaments \cite{Nandakumar2016}. The actual conductance evolution trajectory will depend on factors such as magnitude and duration of programming pulses, the thickness of the dielectric layer, distribution of migrating ionic species and imperfections in the dielectric.   {In our device, the resistivity contrast between the filament region and the dielectric is supposedly reduced via the doping of Cu atoms during annealing. In the blue curve in Fig.\ref{fig2}a, along the lower right arrow, we observe that the device resistance starts to decrease around 250\,mV and an abrupt jump is observed only just before the device conductance hit the $G=G_0$ line indicating the formation of a full filament connecting the top and bottom electrode. While the actual conductance modulation started at around 250\,mV, the atomic rearrangement within the dielectric formed a full filament only by around 450\,mV.  The orange curve in Fig.\ref{fig2}a indicates that by limiting the sweep voltages to lower amplitudes, intermediate states can be obtained.} A negative voltage of sufficient magnitude reverses the ionic migration direction, reducing the extent of the metallic filament and bringing the device back to the HRS states (RESET transition). This bipolar switching behavior indicates that the conductance transitions in the device are due to the field-driven motion of charged ions through the dielectric.

\subsection{Demonstration of spike timing dependent plasticity}

Our demonstration assumes that the device is connected between two spiking neurons.
To emulate the spike timing dependent plasticity in our devices, the programming waveforms are designed as shown in Fig.\ref{fig1}c.  The spikes from the pre- and post-synaptic neurons are converted to waveforms ($V_{pre}$ and $V_{post}$) mimicking the action potential in the biological neurons. These waveforms are applied at the instants of spike activity of the two neurons to the respective terminals of the synaptic device.  The waveforms, $V_{pre}$ and $V_{post}$, are constructed using two decaying exponentials (equation (\ref{eq1})) to capture the depolarization-repolarization-hyperpolarization cycles in the biological action potential waveforms.

\begin{align}
\label{eq1}
\begin{split}
V_{pre}(t)&=A_1e^{-t/\tau_m}u(t)-A_2e^{-(t-3\tau_m)/\tau_s}u(t-3\tau_m)\\
V_{post}(t)&=A_2e^{-t/\tau_m}u(t)-A_1e^{ -(t-3\tau_m)/\tau_s}u(t-3\tau_m)
\end{split}
\end{align}
where $A_1=0.1\,$V, $A_2=0.25\,$V, $\tau_m=3\,$ms, $\tau_s=30\,$ms and $u(t)$ is the Heaviside step function. The amplitudes of  V$_{pre}$ and V$_{post}$ are chosen such that they are below the minimum SET and RESET voltages of the device when there is no or very little overlap between the waveforms. However, when the spikes are closer, the magnitude of the instantaneous voltage across the device increases, resulting in non-volatile conductivity modulation.  In our experiment, the time constants of the programming waveforms are chosen to have non-zero overlap between the pre- and post-synaptic neuron spike waves for a desired duration for this STDP window. Our programming scheme assumes that the $V_{post}$ signal is sent in the backward direction from the post-synaptic neuron when it spikes. If spikes from both the pre- and post-synaptic neurons occur close in time, there will be significant overlap in the voltage waveforms generated across the device. Depending on the direction and magnitude of the field across the device, the conductance of the device could increase or decrease.

For the experimental demonstration, we assume that the Cu electrode is connected to the post-synaptic neuron and W to the pre-synaptic neuron. However, instead of applying the $V_{pre}$ and $V_{post}$ to the respective terminals we compute the difference waveform $V_{post} - V_{pre}$ as a function of time and is applied to the Cu terminal with the W electrode at ground (Fig.\ref{fig2}b). Such waveforms were created using an arbitrary waveform generator for different spike time differences. Example waveforms applied to the device for a positive and negative time difference of $5\,$ms and the corresponding device conductance evolutions are plotted in Fig.\ref{fig2}c. Each programming waveform is appended with an initial and final non-disruptive read pulse of $50\,$mV to measure the device conductance. As indicated in the figure, the $\Delta t = 5\,$ms signal leads to potentiation and $\Delta t = - 5\,$ms leads to depression in device conductance. The average conductance change for the spike time differences in an interval of $[-40\,$ms, $+40\,$ms$]$  is plotted in Fig.\ref{fig2}d based on 400 randomly chosen $\Delta t$s. The average   $\Delta G_{norm}$ ($=  ({G_f-G_i})/{min(G_i, G_f)}$) versus $\Delta t$, where $G_i$ is the initial and $G_f$ is the final conductance for a spike time difference of $\Delta t$, is similar to the STDP response from a biological synapse shown in  Fig.\ref{fig1}b. In Fig.\ref{fig2}e we show the conductance evolution of the device during the STDP measurement. We observed that the device always stayed below the quantized conductance level of $G_0(=2e^2/h)$ with its minimum conductance at $0.016\,G_0$. The filamentary paths often formed in the conductance-bridge resistive memory devices act as nanoscale electron channels and result in conductance levels which are integer multiples of $G_0$. The sub-quantized levels in our device are indicative of non-filamentary atomic rearrangement based conductance modulation. Further, the superimposed orange curve, which is a cumulative representation of the sign of the applied $\Delta t$s during each spike-time-difference based programming event follows the same trend as the device conductance evolution. Thus, the device plasticity has successfully captured the overall causal-anti-causal spike pair relations. In Fig.\ref{fig2}f we plot the  distribution of energy consumed in the synaptic device during the propagation of $ V_{pre} $ and $ V_{post} $ spike waveforms as a function of the time difference between them.  The average energy consumed per spike is $5\,$nJ based on the spike triggered waveforms and the average device conductance range; the energy consumed slightly increases or decreases depending on whether the programming event leads to  potentiation or depression of the synapse.

\subsection{Phenomenological model for state-dependent conductance update}

\begin{figure}[!ht]
	\centering
	\includegraphics[scale=0.5]{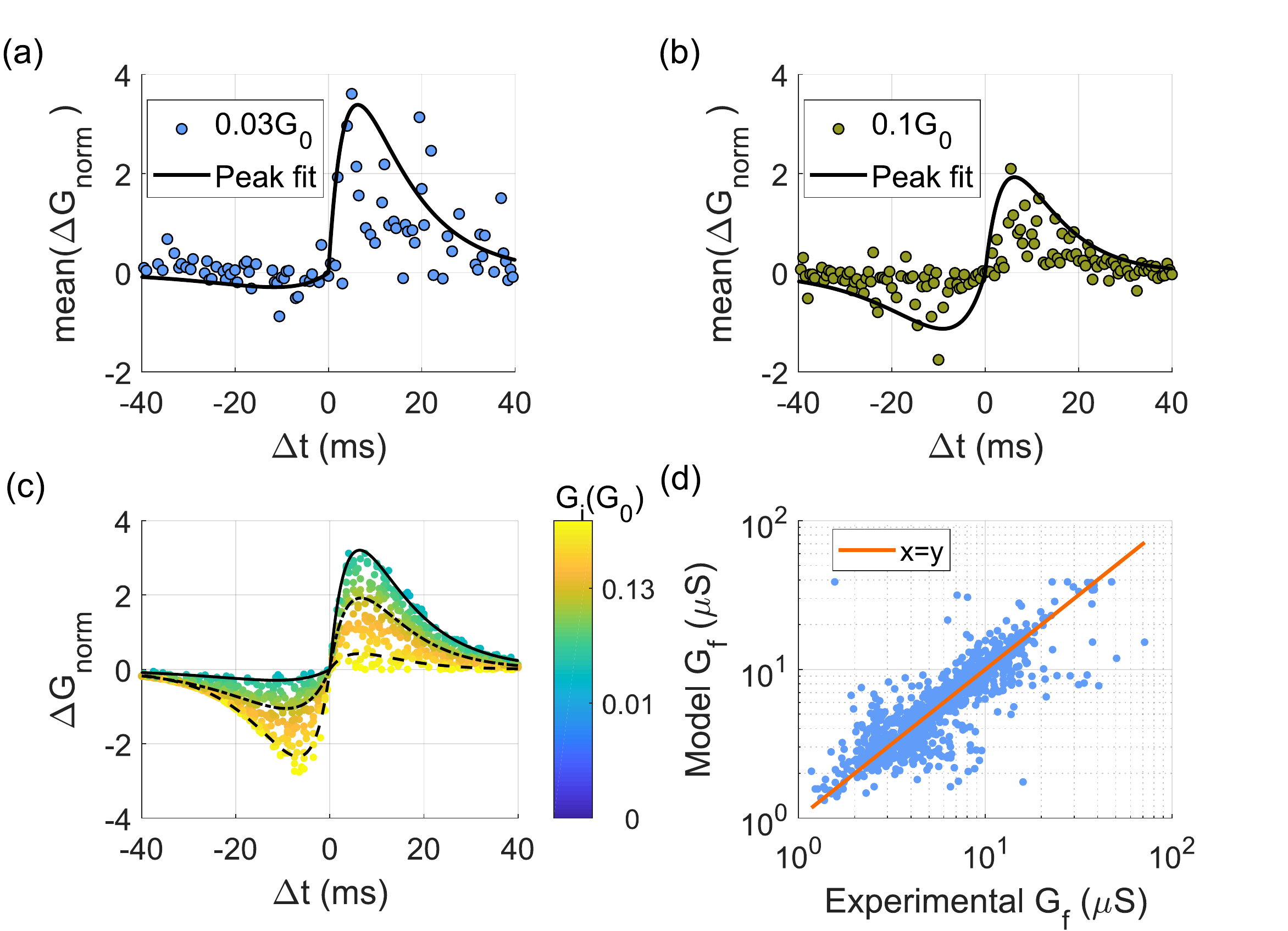}
	\caption{\textbf{Device state-dependency and  phenomenological model} (a) Average $\Delta G_{norm}$  defined as $(G_f-G_i)/min(G_i, G_f)$ response from the device measurement when the initial conductance  is below $0.05\,G_0$ with an average of $0.03\,G_0$  and when (b) it is in the range  between $0.05\,G_0$ and $0.16\,G_0$  with an average of $0.1\,G_0$   (c) The response of the  phenomenological model, when programmed with a  sequence of randomly chosen $\Delta t$s. (d) The device conductance after the application of the pulse ($G_f$),  calculated from the phenomenological model is well correlated with the experimental values ($R^2\sim 0.56$), for the same initial conductance values and programming $\Delta t$s as in the experiment  over a dynamic range of two orders of magnitude.}
	\label{fig3}
\end{figure}

A careful analysis of the device response reveals the state dependency of the conductance change as a function of the timing difference of the applied waveforms. We study the device response for three regimes, demarcated in units of quantum conductance $G_0$: (a) when $G_i< 0.05\,G_0$, (b) when $0.05\,G_0<G_i<0.16\,G_0$, (c) when $G_i>0.16\,G_0$. From the average $\Delta G_{norm}$ vs $\Delta t$ plotted for different ranges of initial conductance, we observe that when the initial conductance is in the intermediate range, conductance change in the direction of potentiation and depression is in the same range for the same spike time difference, while potentiation is pronounced and depression is weak in the low initial conductance regime (Fig.\ref{fig3}a,b)).  Albeit noisy, a similar trend was also visible in high initial conductance regime, where device shows more tendency for conductance depression than potentiation. This behavior indicates a reduction in the incremental conductance change and a conductance saturation as the device reaches closer to its upper and lower conductance limits.

To study how such device characteristics will be effective in emulating synapses in neural network implementations, we developed a computationally simpler phenomenological model that predicts the next state of the device, $G_f$, given the current state, $G_i$, and spike pair time-difference, $\Delta t$.  Our model essentially captures these state dependent conductivity modulation characteristics by modelling the normalized change in conductivity $\Delta G_{norm}$ as:\\
When $\Delta t >0 $, 
\begin{equation}
\Delta G_{norm}  = A\exp\left(\frac{-\Delta t}{\alpha_{ap} + g \beta_{ap}  } \right) - A\exp\left(\frac{-\Delta t}{\alpha_{bp}+ g \beta_{bp}  } \right) 
\label{eq_pot}
\end{equation}
and  when $\Delta t \leq 0$,
\begin{equation}
\Delta G_{norm}  = -A \exp\left(\frac{\Delta t}{\alpha_{an}+ g \beta_{an} }\right) +A \exp\left(\frac{\Delta t}{\alpha_{bn}+ g\beta_{bn}  }  \right)
\label{eq_dep}
\end{equation} 
$A=9$, $g = \log_{10} (G_i/G_{0})$, and other  parameters are listed in Table \ref{params}. Further, the model is limited to operate within a conductance range of $G_{min} = 0.016\,G_0$ and $G_{max} = 0.5\,G_0$. The equations (\ref{eq_pot},\ref{eq_dep}) are formed as the difference of two decaying exponentials with different time constants (see appendix \ref{app:datafit} and Supplementary note 2). The time constants are functions of the device conductance and converges to the same value at the boundary points such that the conductance change gradually becomes zero. 
Such state dependent behavior is akin to the saturating conductance responses observed in many of the gradual conductance and STDP demonstrations in the memristive devices\cite{jackson13, Prezioso15, Wang2016, Chen2016,Park2013b,Prezioso2016, Pedretti2017}.

\begin{table}[ht]
	\caption{Parameters used in the phenomenological model}
	\centering    \begin{tabular}{cccc  }
		\hline
		Symbol&Value  &Symbol&Value  \\
		\hline
		
		$\alpha_{ap}$   & $5.2\,$ms                  &$\beta_{ap}$            & $-3.8\,$ms \\            $\alpha_{bp}$   & $6.9\,$ms                  &$\beta_{bp}$            & $1.9\,$ms \\    
		$\alpha_{an}$   & $9.1\,$ms                  &$\beta_{an}$            & $-1.9\,$ms \\            
		$\alpha_{bn}$   & $2.3\,$ms                  &$\beta_{bn}$            & $-5.7\,$ms \\\hline    
	\end{tabular}
	\label{params}
\end{table}

The STDP response from the phenomenological model is shown in Fig.\ref{fig3}c, where the model conductance is initialized to $1\,\mu$S and is programmed with a sequence of random $\Delta t$s. To compare the model and device response, the model is initialized with the exact device conductance measured before each $\Delta t$ from the experiment and the correlation between the model and the device conductance responses after each $\Delta t$ based STDP programming is plotted in Fig.\ref{fig3}d.    The $R^2$ estimation between these experimental and model conductance values measured over two orders of magnitude is $ 0.56$. 

\subsection{Supervised learning emulation}

Next, we discuss how this device could be used in an exemplary spiking neural network (SNN) for implementing event-triggered learning. An STDP derived supervised learning algorithm, similar to the ReSuMe\cite{Ponulak10resume}, is used to train a network with 1000 inputs neurons and one leaky-integrate and fire (LIF) output neuron  (Fig.\ref{fig4}a) (see appendix \ref{app:snn_emulation} for details on SNN simulation). The task is to determine the weights of the 1000 synapses of the output neuron such that it creates spikes at the desired instants as dictated by a teacher neuron when they are excited by spike streams generated by a Poisson process (Fig.\ref{fig4}c,d). The phenomenological model for the device plasticity was employed to emulate the synaptic behavior during the training of the SNN. At the beginning of training, synapses are initialized {as a distribution around the geometric mean of the maximum and minimum conductance of the model ($G_{ref} = \sqrt{G_{max} G_{min}}$).  This $G_{ref}$ is considered as a reference level around which the synaptic weights are allowed to vary during the training such that individual synapses could be either excitatory or inhibitory. Implementation of such a reference level in hardware may require an additional memory device along with each synaptic device (Supplementary note 3). The training rule for the synapses is shown in Fig.\ref{fig4}a,b. When the teacher neuron spikes, the synapses are potentiated based on the time elapsed since the most recent input spike. Similarly, the synapse will be depressed when there is an observed spike from the output neuron, based on the time difference with the last input spike. When the output neuron spike coincides with a teacher neuron spike (i.e., the desired response is obtained from the network), there will not be any synaptic modulation. The amount of potentiation or depression for each time difference is determined in a state-dependent manner using the device STDP model. 
{The input and desired spike patterns are presented to the network and the synaptic conductance update process is repeated during each training epoch. The evolution of the spikes observed from the output neuron as a function of the training epochs is shown in Fig.\ref{fig4}d. The network generates all the spikes at the desired times within $\pm 10\,$ms in 9 epochs. The conductance evolution of a few synapses from the SNN during the course of the training is shown in Fig.\ref{fig4}e.} 
	\begin{figure}[!ht]
		\centering
		\includegraphics[width =0.97\textwidth]{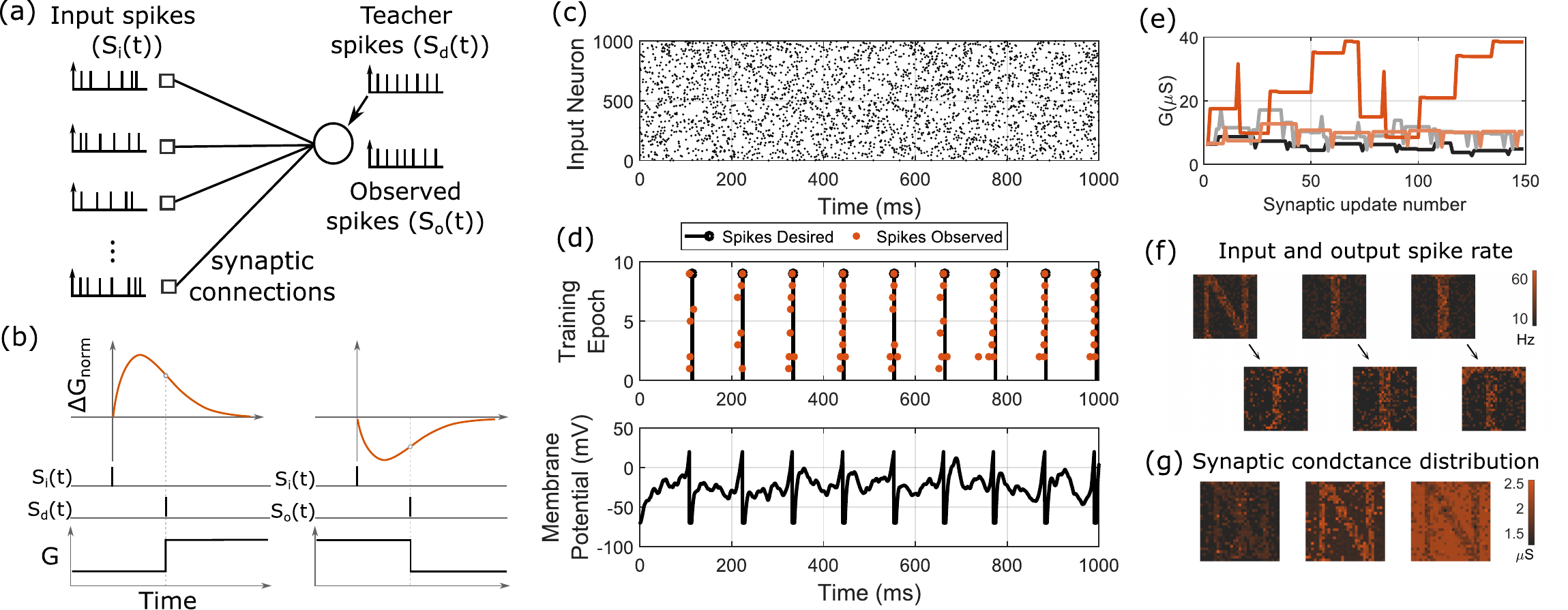}
		\caption{\textbf{Supervised training of SNN} (a) An exemplary spiking neural network with $N$ input neurons connected to a single output neuron. The training task is to discover the synaptic weights such that the output spike response, $S_o(t)$, matches the desired response, $S_d(t)$, for a specified input spike excitation, $S_i(t)$.  (b) Training rule: the synapse is potentiated or depressed based on the time difference of desired or observed spikes respectively from the efferent spike using the STDP model. (c) A raster plot of the spike streams from each input neuron in a $1000 \times 1$ SNN is shown. (d) The desired and observed spike trains over the training epochs (top) and the final membrane potential (bottom) from the output neuron as a function of time. (e) Device conductance evolution for four exemplary synapses during the training of the $1000 \times 1$ neural network is shown.  {Each synapse shows a potentiation at the times of desired  spikes if it has received an input spike recently. Similarly, the synapse shows depression at the times of observed output spikes if the synapse has received an input spike recently.} (f) Input (top) and observed output (bottom) neuron spike rate in an SNN trained for sequence prediction. (g)  Relative conductance distribution of synaptic device models connected to three output neurons showing the features learned after training. The first set of synapses is connected to a neuron responding to letters N and J. The second set of synapses is connected to an output neuron responding only to letter N. The third output neuron had a desired spike having anti-causal relation with most of the input spikes leading to an effective synaptic depression.}
		\label{fig4}
	\end{figure}

	Such algorithms are at the heart of supervised learning platforms in spike domain. Once input and output data are translated to spike domain, it can be used to efficiently train networks for different tasks, provided the synaptic realization has sufficient analog programmability. The $N \times 1$ network could be extended for more complex problems. For example, we realize a $900 \times 900$ SNN whose synapses are represented by our phenomenological model (Supplementary Note 4).  The network acts as a sequence predictor for English letters N$\rightarrow$J$\rightarrow$I$\rightarrow$T such that when the input layer is presented with one of the letters, the network creates the image of the next letter in the sequence at the output layer. The pixel intensities of $30 \times 30$ grayscale images are used as rate constants for a Poisson process to generate the corresponding input and desired output spike streams. The network synapses are trained using the same supervised STDP rule as before. The  input and the resulting output   spike rates from the network after 40 epochs of training are mapped to the input and predicted output image as shown in Fig.\ref{fig4}f. The conductance distributions of the synapses connected to three output neurons (akin to a receptive field) are shown in Fig.\ref{fig4}g.
  {Here, the first set shows the conductance distribution of synapses connected to an output neuron that should spike when either an N or J is present at the input. The second set of synapses corresponds to a neuron that should spike only when an N is present at the input. The third set of synapses corresponds to a neuron that should not spike when an N is present at the input. These learned conductance distributions illustrate the ability of the nano-scale device to capture essential features to perform the desired task under the constraints of limited dynamic range and state-dependent conductance update. }
	\subsection{Unsupervised learning emulation}
	
	We now evaluate the unsupervised learning capability of device based on the  observed  STDP characteristics  on the commonly used benchmark task of classifying handwritten digits from the MNIST dataset. We study two fully connected SNNs - both the networks receives 784 inputs (corresponding to $28\times28$ pixels in the image); the first network only has  10 LIF output neurons, while the second has 30 LIF output neurons and in both the SNNs a winner-take-all dynamic is implemented between the output neurons (Fig.\ref{fig5}a).  The unsupervised classification performance has been shown to improve by increasing the number of output neurons \cite{Querlioz2011, Diehl2015}. The  $ 28\times 28 $ pixel intensities of the training images were binarized and presented as spike streams to the input of the SNNs.  Each image is applied for $200\,$ms. \textit{Off} pixel values do  not receive any spikes and \textit{on} pixel values  receive a spike at 50\,ms. Corresponding to each spike, a voltage signal $ -V_{pre}(t) $ was applied to the synapses. The LIF neurons integrate the currents to generate output spikes and the synaptic conductance values are modulated based on the STDP model. The  winner-take-all dynamics in the output layer resets the integrated membrane potential of all the output neurons  when any one of the neuron issues a spike, preventing them from spiking for the next $3\,$ms. For each output neuron spike, the synapses with a pre-neuron spike within a $40\,$ms duration are potentiated based on the spike time difference and those without a spike is depressed assuming a spike time difference of $-60\,$ms. Further, we maintained homeostasis between the output neurons by adjusting the threshold voltage of the LIF neurons every $100$ images such that all the output neurons have a similar spike rate on average \cite{Querlioz2011}.

	We emulated the synapses with 1, 5, and 10 devices as a method to improve the conductance change granularity \cite{Boybat2017}. A reference corresponding to the minimum conuctance level was assumed per synapse such that synaptic conductance is $ n(G - G_{min}) $ where $ n = 1,5,$ or 10. To implement pulse overlapping STDP programming in a multi-memristive synapse, as envisioned in our conductance update scheme, the output neurons integrate current from all the devices in a synapse, however, the $ V_{post} $ is sent back through only one of the devices chosen cyclically. In a crossbar array of memristive devices, this will involve a neuron integrating currents from multiple bit lines and sending a voltage through one of the bit lines when the neuron issues a spike. This permits uniformly distributing the programming events across multiple devices in a synapse.  
	
	Further, to evaluate the learning performance in the presence of programming noise, weight updates are assumed to be Gaussian random processes such that the mean of the normalized conductance updates ($ \mu $) are determined using the equations~\ref{eq_pot} and \ref{eq_dep} and standard deviations $ \sigma $ are determined as a fraction (varied from 0 to 0.5) of  $ \mu $. Cumulative conductance evolution when the model was subjected to a sequence of programming pulse corresponding to spike pairs of $ |\Delta t|= 5\,$ms and $ 15\,$ms is shown in Fig.\ref{fig5}b. 
	
	During each training epoch, the 60,000 images in the database were presented to the SNN and the weights were updated using STDP for 10 epochs. The features learned by the weights connected to ten of the output neurons at the end of training  is illustrated in Fig.\ref{fig5}c. We observe that the average features corresponding to different digits in the dataset are captured by the weights corresponding to different output neurons. The image labels were not used to determine the weight updates. At the end of every epoch, each output neuron is assigned a label corresponding to the digit for which it generated the maximum number of spikes during the presentation of the last 10,000 images from the training set. These labels are used to determine the prediction performance of the SNN on a test set of 10,000 images which were not shown  during training (Fig.\ref{fig5}d). Under the memristive STDP response, we observe the classification performance to improve with the number of output neurons and with more number of devices per synapse. More remarkably, the test accuracies seem extremely tolerant to the programming noise. We achieved a maximum test set accuracy of {76.65\%} using 30 output neurons and 64.2\% using 10 output neurons, both of which have 10 memristive devices per synapse and were programmed with $ \sigma/\mu(\Delta G_{norm}) = 0.5 $.


		\begin{figure}[!ht]
		\centering
		\includegraphics[width =0.97\textwidth]{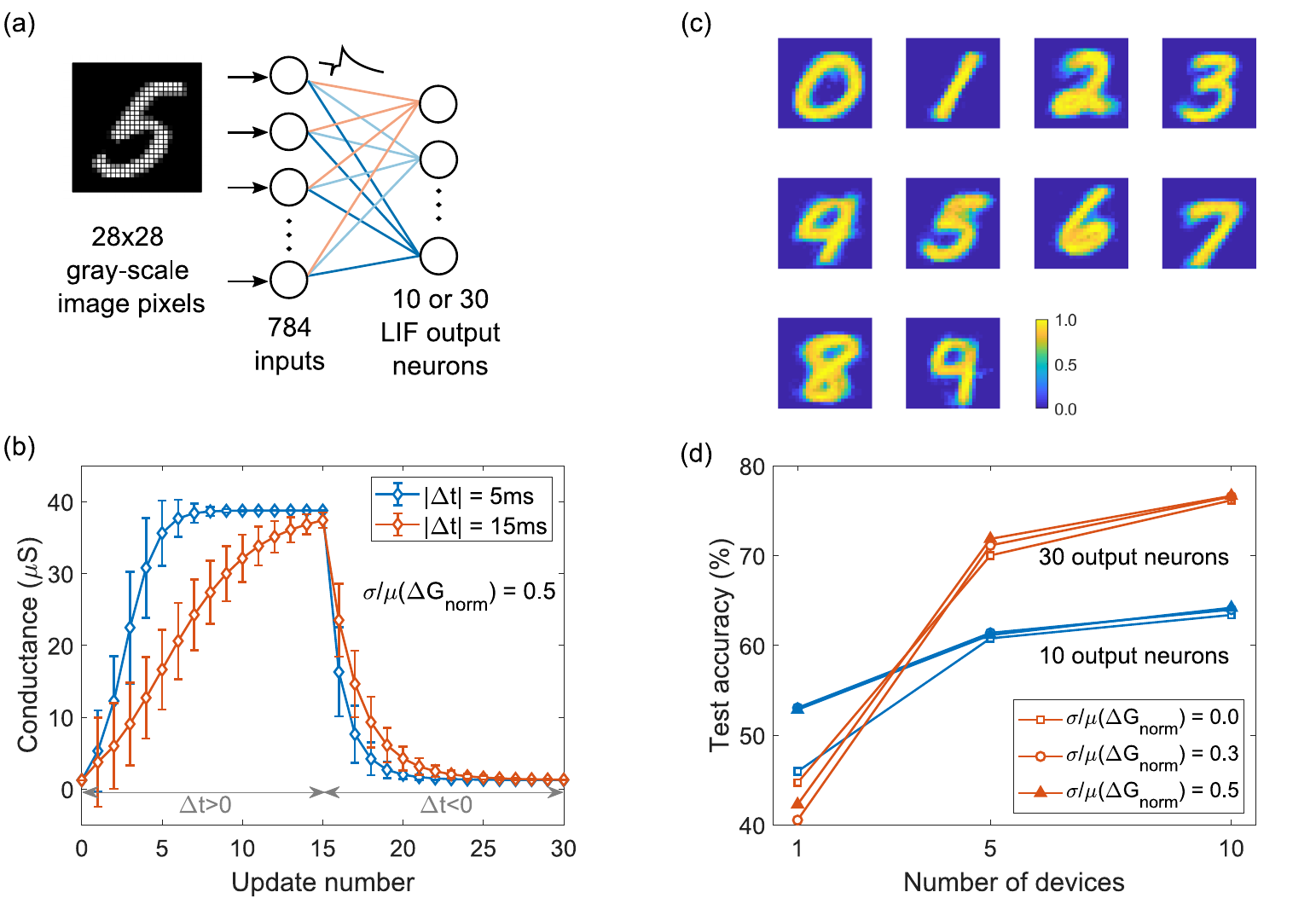}
		\caption{\textbf{Unsupervised training of SNN}(a) SNN used for the unsupervised training of handwritten images of digits. (b) The conductance evolution behavior of the phenomenological model used to emulate the SNN synapses. Model response when trained with a sequence of programming pulses corresponding to spike time intervals of 5\,ms and 15\,ms is are shown. (c) Features learned by the weights (normalized in the figure) corresponding to ten output  neurons. (d) The classification performance of the SNN on 10,000 test images from the MNIST dataset. The results are average of three random simulations. The performance corresponding to different amount of programming noise added to the synapse model during training is shown.}
		\label{fig5} 
		\end{figure}



	\section{Discussion}
	
	While there have been numerous demonstrations of CBRAM devices, including some based on the Cu/SiO$_2$/W material system, they were non-volatile memory devices with binary states and high on-off ratios ($\sim 10^3$) \cite{Kozicki2005, Schindler07,Russo2009, Valov2011, Jameson2013, Jameson2016a, Gonzales2016}. For neuromorphic systems that learn and adapt, it is desirable to have synaptic devices with incremental conductance changes. STDP behavior has also been demonstrated before in silicon based \cite{Jo2010} and oxide based memristive devices  \cite{Park2012a,Rajendran18, Prezioso15, Covi2016}. However, these devices are characterized by either high operating voltages (typically $>$ 1\,V) or high operating current (0.1\,mA to 10\,mA) or low on-off ratio ($<\,$10). Also, gradual conductance change by modulating the filament thickness in CBRAM \cite{Mahalanabis2016, Lim2018} is undesirable due to the high power consumption associated with transport through metallic filaments \cite{Jameson2016a}.
	In contrast to the bipolar memristive device which implements STDP via overlapping programming pulses, unipolar devices such as phase-change memory suffer from the requirement for more complicated programming waveforms or circuits to implement STDP  behavior\cite{jackson13}. Hence, our demonstration of STDP in the sub-quantized (conductance $ < 38.7\,\mu $S) regime in a CMOS compatible memristor  operating below 500\,mV is a significant step towards realizing efficient neuromorphic learning systems. 
	
	The unsupervised learning performance using the state-dependent STDP model of our device on the standard MNIST benchmark dataset is at par with similar results in literature. For instance, approximately 60\% accuracy in an SNN with 10 output neurons and 80\% accuracy with 50 output neurons have been reported using synaptic models with tunable update granularity or learning rate \cite{Querlioz2011, Diehl2015}. However, one of the major challenge in synaptic device based learning systems is the fixed range of realizable conductance changes. Multi-memristive  synaptic architecture \cite{Boybat2017} and mixed-precision architecture   for supervised learning \cite{Nandakumar2018} have been proposed to compensate the limited device granularity. Our demonstration of software-equivalent  performance for unsupervised learning using realistic device STDP models, and its high tolerance to programming noise during training, is thus a promising step towards realizing functional learning machines using extremely scaled stochastic analog memory devices.

	Such analog memory devices can be integrated in crossbar arrays to create high-density synaptic networks. While  the sneak path issue is not so severe during parallel weighted summation of the neuronal inputs, the necessity to perform device programming without altering the neighboring devices may make a selector device at the cross point in series with the synaptic device desirable. The bipolar nature of the conductance update in the memristive device permits STDP programming by overlapping pulses from the bit lines and word lines of a crossbar array. Several programming methods have been proposed to achieve this, including the amplitude modulated rectangular pulse sequence schemes \cite{Yu2010b} and those based on continuous analog waveforms \cite{ZRam2011, Rajendran18} drawing different levels of biological inspiration. One of the initial approaches has been to convert the sign and magnitude of the spike-time difference into the polarity and width of a rectangular programming pulse \cite{Jo2010}. However, this necessitates a global circuit to generate the programming pulse with tunable width. In a true pulse overlapping programming scheme, the main challenge is the difference in the device switching time and the desired STDP window. The exponential programming waveforms we used for the CBRAM has been tuned such that sufficient electric field is set up across the device to initiate an ionic drift while keeping the duration and amplitude of the voltage sufficiently low to avoid rapid switching transitions to the extreme conductance levels. This is due to the fact that the ionic migration velocity in the device dielectric is exponentially dependent on the electric field and hence the actual conductance evolution characteristic is also dependent on the time-scale and shape of the programming waveform \cite{mottgurney1948, Schindler2009, Menzel2015}.  If the STDP based learning is expected to detect correlation between events in real time encoded using sparse spikes, the STDP window, and correspondingly the input waveforms, need to last few tens of milliseconds, while the switching time of the CBRAM devices has been shown to vary from {tens of microseconds to a few milliseconds} \cite{Russo2009}. 
	The long programming waveforms also lead to high energy consumption in the synapses.  The average energy consumed per spike in the memristive device is approximately 5\,nJ (which translate to an average of 10\,nJ per spike pair used for STDP programming). Based on the average spike rate received by a synapse and the number of synapses in the SNN, the 
	energy consumption in the synaptic network can be estimated. For example, approximately 0.1\,mJ per epoch is required in the $ 1000\times 1 $ analog memory array for the supervised training and 0.16\,mJ per image in a $ 784\times 10 $ analog memory array for the unsupervised learning. To realize large scale SNNs approaching the efficiency of the brain, further optimization of these synaptic devices is required, including operating in lower conductance ranges, switching in smaller time scales, and innovative STDP programming schemes that take into account the device programming characteristics and the application at hand.

	STDP like local learning rules are key to the decentralized and parallel processing capabilities of the biological neural networks. Realization of biological plasticity mechanisms in nano-scale memristive devices when combined with their high integration density and scaling potential enables power efficient implementations of large-scale learning networks. While this proof-of-concept demonstration establishes the basic feasibility of our device for learning, there are challenges in terms of device reliability, and variability that warrants further research and optimization. For example, the Cu/SiO$_2$/W based devices we fabricated was responsive to approximately 1000 STDP measurements. The higher mobility of Cu in SiO$_2$, though useful for low voltage operation, might be the reason behind the rapid deterioration in endurance and retention performance\cite{Nandakumar2016}. Further, the stochastic nature of atomic rearrangement upon programming results in stochasticity in the conductance modulations as well. Also, the device-to-device and cycle-to-cycle variation of the resistive memory devices could affect the array-level performance, though it is expected that some of these reliability issues could be mitigated by improved industrial fabrication processes. Furthermore, {linearity and precision of conductance update of such nanoscale devices are also traits that need to be improved for its applicability to a wider class of training applications. Meanwhile,}   the targeted neuromorphic applications are expected to be more error tolerant as decisions are based on overall synaptic distributions   rather than  the absolute conductance levels of any particular device,  making resistive memory based synapses still attractive for cognitive hardware.

	\section{Conclusions}
	We have experimentally demonstrated STDP behavior, an unsupervised local neural learning strategy observed in nature, in a CMOS compatible $500\,$mV memristive device using biomimetic programming waveforms. Using a phenomenological model of the observed characteristics, we have also demonstrated the suitability of the device to implement complex supervised learning algorithms to generate spikes at precise time instances and to perform unsupervised learning on handwritten digits using MNIST dataset. The computationally simple non-linear model helps to study the suitability of such devices in different learning scenarios. With further optimizations, the cyclability and retention characteristics of the device can be improved, enabling the fabrication of crossbar array platforms for large spiking neural circuits for accelerating learning applications.

{
	\renewcommand{\thesubsection}{\Alph{subsection}}

	\section*{Appendices}

	\subsection{STDP programming and plot}
	In this section, we describe how the STDP characterization of the device was performed. We used Agilent B1500 semiconductor parameter analyzer with a B1530 unit. B1530 is an arbitrary waveform generator and fast measurement unit (WGFMU) capable of applying $100\,$ns pulses and accurate current measurement with up to $5\,$ns sampling rate. The set-up is connected to a probe station where the device is probed and characterized. The programming waveforms were created in Matlab and the B1530 was controlled from Matlab using custom functions.
	
	The spike programming waveforms were designed using Matlab and were converted to voltage signals using the WGFMU.  A slow dual voltage sweep with a ramp rate of approximately $2\,$V/s was used to characterize the device for its discrete switching behavior. For the STDP characterization of the device, $1\,$s long patterns where used at a time, and in that duration each $250\,$ms was used for a waveform corresponding to one spike pair. The waveforms were created from data-points at a resolution of $0.5\,$ms.
	Read pulses of  $5\,$ms duration and $50\,$mV amplitude were inserted at the ends and in between the programming waveforms to determine the state of the device. Programming waveforms corresponding to randomly chosen 2000 $\Delta t$s ($\in [-80\,$ms,$+80\,$ms$]$) were applied to the Cu terminal of the device. The resulting current as a function of time is read using the measurement unit.  The average current during the read pulse is divided with the read voltage to get the device conductance changes due to each STDP event. However, after approximately 1000 programming events the device became relatively unresponsive and was stuck to a narrow range of conductance. Also, the conductance response outside the window of $[-40 $ms, $+40$ms$]$ was not well correlated and are discarded from the study. Now, to determine the STDP plot in Fig.\ref{fig2}d, $\Delta G_{norm}$s corresponding to each unique $\Delta t$ were averaged. For the state-dependent responses in Fig.\ref{fig3}a,b the $\Delta G_{norm}$s were averaged whose $G_i$ were in the chosen range.

	\subsection{Data fitting and phenomenological model}\label{app:datafit}
	Here, we describe how the parameters in the phenomenological model (equation (\ref{eq_pot}, \ref{eq_dep})) are obtained from the device STDP response. 
	We first obtained peak-fits of the average $\Delta G_{norm}$  data points for medium and low initial conductance range separately for the positive and negative $\Delta t$ range (Figure S2 in the Supplementary note 2). The peak-fit curves were approximated by functions of the form $f = A(exp(-\Delta t/\tau _A) - exp(-\Delta t/\tau _B))$. The parameters $\tau_A,  \tau_B$ and $A$ are obtained by minimizing the error with the peak-fit points using gradient descent.  Next, these $\tau_A$ and $\tau_B$  are approximated as linear functions of $log(G_i)$ as $\tau = \alpha+ \beta\times log(G_i/G_0)$, where $G_0 = 2e^2/h$. For the conductance potentiation, these lines for $\tau_A$ and $\tau_B$ meet at a higher conductance and for depression, they meet at a lower conductance.  The $G_i$ values at which these lines meet define the boundary conductance points for the STDP model. For our model fitted with the device characteristics, these boundary points are approximately at $0.016\,G_0$ and $0.5\,G_0$ (Fig. S3 in the Supplementary note 2).

	\subsection{Spiking neural network emulation}\label{app:snn_emulation}
	In our SNNs the input layer encodes information in the form of spikes using a Poisson process. An input, such as image pixel intensity, is used as the rate constant for the random process to generate a sequence of events representing the spikes. These spikes are propagated via the synaptic junctions and the resulting currents are integrated by the next layer neurons.  Here, we used LIF models to emulate the neurons where the neurons are leaky capacitors integrating the input current and fire a spike when the integrated voltage crosses a threshold ($\theta$) and then the voltage across the capacitor is reset to neuron resting potential ($E_L$). This LIF neuron dynamics are emulated by the integral, $V_m = (1/C_m)\int \Sigma I_{syn}-g_L(V_m-E_L) dt$ whose approximate solution is determined using second-order Runge-Kutta method.  The capacitor voltage, $V_m$, which emulates the biological neuron membrane potential, is held at $E_L$ for a refractory period of $5\,$ms after a spike.   The $I_{syn}$ is the current through the synapse in response to a spike and is modeled as $W(exp(-t/\tau_1)-exp(-t/\tau_2))$, where $W$ represent the synaptic strength and the difference of two exponentials model the synaptic current kernel. Each spike from the pre-synaptic layer will cause a current $I_{syn}$ to pass through the synapse, and the currents from all such input synapses will be integrated at the post-synaptic neuron. In our simulation we used 
	$\tau_1 = 5\,$ms and $\tau_2=1.25\,$ms,  $C_m = 300\,$pF, $g_L = 30\,$nS, $\theta = 20\,$mV, and $E_L = -70\,$mV and the simulation time-step was $0.1\,$ms. 

}

\section*{Acknowledgements}
The device fabrication and characterization was carried out at IIT Bombay Nanofabrication Facility with partial support from
grant 5102-1 from Indo-French Centre for the Promotion of Advanced Research (CEFIPRA). The analysis, modeling, and training simulations were carried out while the authors were at New Jersey Institute of Technology, Newark, USA.

\section*{Declaration of interest }
The authors declare no competing interests.

\section*{Supplementary material}
Supplementary information on device fabrication, phenomenological model, crossbar realization of synaptic array, and the sequence predictor network is available.


%

\end{document}